\documentclass[useAMS,usenatbib]{mn2e}
\usepackage{graphicx}

\title[Comparison of sunspot group tilt angles]{Comparison of Debrecen and Mount Wilson/Kodaikanal sunspot group tilt angles {\bf and the Joy's law }}
\author[T. Baranyi]{T.~Baranyi\thanks{E-mail: baranyi.tunde@csfk.mta.hu} 
  \\
  Heliophysical Observatory, Research Centre for Astronomy and Earth Sciences, Hungarian Academy of Sciences, \\
  Debrecen, P.O.Box 30, H-4010, Hungary\\
}

\begin{document}

\date{Accepted 2014 December 03; Received 2014 December 01; in original form 2014 Sep 30}

\pagerange{\pageref{firstpage}--\pageref{lastpage}} \pubyear{2014}

\maketitle

\label{firstpage}

\begin{abstract}
The study of active region tilt angles and their variations in different time scales plays an important role in revealing the subsurface dynamics of magnetic flux ropes and in understanding the dynamo mechanism. In order to reveal the exact characteristics of tilt angles, precise long-term tilt angle data bases are needed. However, there are only a few different data sets at present, which are difficult to be compared and cross-calibrate because of their substantial deviations. In this paper, we describe new tilt angle data bases derived from the Debrecen Photoheliographic Data ($DPD$) (1974--) and from the SOHO/MDI-Debrecen Data ($SDD$) (1996-2010) sunspot catalogues. We compare them with the traditional sunspot group tilt angle data bases of Mount Wilson Observatory (1917-85) and Kodaikanal Solar Observatory (1906-87) and we analyse the deviations. Various methods and filters are investigated which may improve the sample of data and may help deriving better results based on combined data. As a demonstration of the enhanced quality of the improved data set a refined diagram of the Joy's law is presented.

\end{abstract}


\section{Introduction}

It has been known for a long time that the line connecting the leading and following portions of a bipolar sunspot group usually tilts with respect to the solar equator \citep{Hale19}. The tilt angle has been regarded as an important parameter of the magnetic field since the publication of the solar dynamo models by \citet{B61} and \citet{L69}. The longest available tilt angle data bases are derived from white-light photographic observations taken at Mount Wilson Observatory ($MW$) in 1917-1985 and Kodaikanal Solar Observatory ($KK$) in 1906-1987 \citep{HGG84, H91b, SGH93}. The magnetograms available since 1974 also allow the study of tilt angle of active regions \citep{WS89, H91a}. Numerous characteristics of tilt angles and their changes have been investigated based on these data bases (e. g. \citealt {H93, H96a, H96b, SGH99, HSG00}). Several papers  have also confirmed that the average tilt angle increases with increasing latitude known as Joy's law reported  by \citet{Hale19} at first.

Recently new methods and data (e.g. tilt angles derived from space-based magnetograms) have been included into the studies. For example, \citet{Dasi10, Dasi13} initiated the study of cycle-to-cycle dependence of the average tilt angles. 
\citet{LU12} developed a new method to derive tilt angles from MW and SOHO/MDI magnetograms. They started to investigate the median tilt angle instead of the mean tilt angle, and they found a persistent asymmetry between the median tilt angles measured in the northern and southern hemispheres in all latitudinal ranges.  \citet{MN13} also paid attention to the differences between hemispheres and cycles and they suggested a revision of Joy's law equations.  It is widely accepted that the tilt angle is primarily due to the Coriolis force acting on initially untilted rising magnetic flux tubes 
(e.g. \citealt{WS91, DC93}) but some authors (\citealt{SGSGH07, KS08, SK12}) have questioned this explanation arguing that the tilt is already established in the source region inside the Sun.
 \citet{Tea13} revealed a new dynamo pattern by separating the tilt angles of large and small solar bipoles. \citet{Jea14} studied the tilt angle scatter in the framework of Babcock-Leighton dynamo models by using the $MW/KK$ data bases. They concluded that the tilt angle scatter has a significant impact on the variability of the solar cycle strength.
 These examples show that a number of new ideas and questions have arisen concerning the tilt angles recently. To reveal more details and answer the open questions, the studies need extended homogeneous long-term measurements of tilt angles or cross-calibration of various data sets derived with different methods.

The longest sunspot group tilt angle data set available after the termination of $MW/KK$ data bases has been derived recently from $DPD$. The $DPD$ tilt angles can be also compared with the tilt angles derived from the SOHO/MDI-Debrecen Data ($SDD$) catalogue with or without considering the magnetic polarity information. In this study, we analyse and compare sunspot group data available in various data bases ($MW/KK$, $DPD$, $SDD$), and some new data derived for the purpose of comparison presented in this paper. The aim of this study is to reveal the special characteristics of these data bases and to look for various filters that can be used to decrease the differences between them.

\section{Data}

The $DPD$\footnote{http://fenyi.solarobs.unideb.hu} data (Gy\H ori, Baranyi \& Ludm\'any 2011)  are measured on daily white-light full-disk photographic plates, which are mainly taken at Debrecen Heliophysical Observatory and its Gyula Observing Station but, in some cases, the data derived from observations of cooperating ground-based  observatories or from space-borne images.  $DPD$ covers the years starting with 1974 with one observation/day time resolution and it contains the position and (whole and umbral) area data for all sunspot groups and each spot in them.  These data allow deriving tilt angle data from them. By using the umbral data of $DPD$, the umbral tilt angle ($T_{DPDu}$) of the sunspot groups can be determined in a similar way as it was done at $MW/KK$ after measuring only umbral areas. In addition, a new type of tilt angle ($T_{DPDws}$) can be derived from $DPD$ by using the whole spot area (WS) of the spots and pores ($T_{DPDws}$). The calculation of the tilt angle data follows the traditional way in both cases \citep{H91b}:

\begin{equation}
   T=tan^{-1}((B_{f}-B_{l})/(L_{f}-L_{l})/cos(B)*sign(|B_{f}|-|B_{l}|))
\end{equation}
where $B_{f}$ / $B_{l}$ are the area-weighted latitude of the following/leading portion,  $L_{f}$, $L_{l}$ are the area-weighted longitude of the following/leading portion, and  $B$  is the latitude of the centroid of the whole group. 
The separation of the leading and following portions can be derived as
\begin{equation}
   S=\sqrt{(B_{f}-B_{l})^2+((L_{f}-L_{l})cos(B))^2}.
\end{equation}

The $DPD$ does not contain magnetic polarity information on spots but the available magnetograms are frequently taken into account while grouping spots. Thus, the $DPD$ tilt angles based on estimated or measured polarities may be closer to the magnetic tilt data than they would be in the case of automatic grouping based simply on proximity.

The Mount Wilson Solar Observatory has taken full-disk broad-band observations called White Light Directs once per day since 1906. The images for 1917-1985 were measured to determine the sunspot group tilt angles with the method described by \citet{HGG84}. The umbral area and mean position of each spot present within $60^{\circ}$ longitude from the central meridian (LCM) were measured by measuring the position of two successive vertices of the superposed quadrilateral. In the next step, a program based on their proximity grouped the spots. A spot was included in a group if its position was within a distance of $3^{\circ}$ in latitude or $5^{\circ}$ in longitude from another spot in the group. The area-weighted position of the leading (following) portion was derived from the data of umbrae in the portion located to the west (east) of the area-weighted centroid of the sunspot group \citep{H91b}. The same measuring method was applied to the white light images of Kodaikanal Solar Observatory \citep{SGH93} to determine the $KK$ tilt angles of sunspot groups and their daily changes for the year 1906-1987. These traditional sources of tilt angle data ($T_{MW}$ and $T_{KK}$) are available at NOAA/NGDC\footnote{http://www.ngdc.noaa.gov/stp/solar/solardataservices.html}.

The $MW$ Directs was digitized and published in the frame of Mt. Wilson Solar Photographic Archive Digitization Project\footnote{http://ulrich.astro.ucla.edu/MW\_SPADP/index.html} \citep{BUB10}. At present, the $MW$ Photo Archive contains scans of white-light observations for the intervals 1917-18 and 1961-67 overlapping with the $MW$ tilt angle data base. These images can be evaluated with the software developed for $DPD$. We selected the Directs available for 1961-67, and the position and area data of spots were measured with the software of $DPD$. The new tilt angle data set ($T_{MWDPD}$) was created by using the umbral data like in the case of $T_{DPDu}$. 

The Mount Wilson sunspot polarity drawings\footnote{http://obs.astro.ucla.edu/intro.html} (see e.g. \citealt{Tea14}) are often available close in time to the $MW$ Directs. Thus, by comparing the $MW$ drawings and images, we can usually determine or estimate the polarity of the spots involved in the leading and following portions of the groups in $T_{MW}$ and $T_{MWDPD}$. 

The Greenwich Photoheliographic Results ($GPR$) catalogue (1874-1976) contains position and area data of sunspot groups measured in photographic observations taken at Royal Observatory of Greenwich and at a few other observatories. 
It can provide additional information for us on the correct separation of groups and on the nearby sunspot groups within the overlapping interval with the $MW$ data set. We used the digital version of $GPR$ available at NGDC.
Concerning further details on $GPR$ data and their digital versions see e.g. the paper by \citet{Wea13}.

For the time interval 1996-2010, some comparisons can be made by using the $SDD$\footnotemark[1] tilt angle data base \citep{Gyea11} derived from SOHO/MDI quasi-continuum images. 
Four types of tilt angle can be calculated based on this data set but we used only two of them. The $T_{SDDu}$ is the "traditional" tilt angle like $T_{DPDu}$ while in the case of $T_{SDDupo}$ the separation of leading and following portion is based on the polarity information of umbrae derived from MDI magnetograms. 
In the case of $SDD$ and $DPD$, the additional selection criterion of $|LCM|<60^{\circ}$ was applied because the measurements were confined to this region at $MW/KK$.

\section{Direct comparison of $MW$ data}

To reveal the major characteristics of $MW$ data, we compared two tilt angle data sets ($T_{MWDPD}$ and $T_{MW}$) based on the $MW$ Directs. We used the $GPR$ to match the data of these two data set.  In $T_{MWDPD}$, we identified the sunspot groups according to $GPR$. 
We also tried to match the sunspot groups in $T_{MW}$ to the groups in $GPR$ by using their heliographic latitude and LCM data corrected to the time difference of observations. The comparison of the available data after Jan. 1, 1965 revealed that the time of observation of $T_{MW}$ is in UT but the correct date is the next day.  Before 1965, the date of $T_{MW}$ is correct but the published time has to be subtracted from $24^h$ to derive the correct time of observation in UT. After the needed corrections, the data of sunspot groups in $T_{MW}$ and $T_{MWDPD}$ can be matched quite well. (This is also valid for the $KK$ data set because of the same measuring method and software.) 

Identifying the matching cases, we generated the combined data set $T_{MW}\&T_{MWDPD}$ containing 506 sunspot groups with tilt angles for 1961-67. This pilot data set together with the Directs and polarity drawings allows us to determine such characteristics of the measured sunspot groups which are not published in $T_{MW}$. The pilot data are used to extrapolate the result for the whole interval of $T_{MW}$ (and $T_{KK}$) and for the following era. The correctness of the extrapolation is checked by using $T_{SDDu}$ in some cases.

The position data of leading and following portions published in $T_{MW}$ allowed us to estimate which spots were measured as members of the given portions in the Directs. When we compared these portions in the image with their matching parts in the $MW$ polarity drawings, the estimation of the magnetic polarities of spots also became available. In this way, we can determine whether a given portion of a group indicated as leading (following) portion published in $T_{MW}$ corresponds to the magnetic portion of a bipolar group correctly. Table 1 shows that in the most cases, the identification of the portions of bipolar groups was correct in $T_{MW}$.

However, in about 26\% of the cases, the measured tilt angles refer to unipolar groups that contain only spots of one polarity, these data necessarily distort the statistics. The estimated number of these cases based on the $MW$ observations is about the same as the percentage based on the $SDD$ tilt data derived by comparing the number of the traditional tilt angles and that of the tilts with polarity information. The percentage of tilt angle of unipolar groups is probably similar in any tilt angle data base in any time interval.  

The other source of distorted statistics is the false grouping of spots. In the pilot data set, about 14\% of the $T_{MW}$ data refers to such groups which are unipolar because of the measuring method. In these cases, both the leading and following parts of a group are measured but they are separated and assigned as two independent unipolar groups. These divisions are caused by the automatic separating algorithm in those cases when the opposite polarity portions have larger distance than the threshold for grouping.

A further possible source of false tilt data may be the accidental close proximity of nearby groups. By examining the studied sample we can estimate that $\sim$ 4\% of $T_{MW}$ belongs to this type of cases. The percentage of such cases may be probably smaller in $DPD$ because of the use of magnetograms but a few percent of incorrect grouping can also be estimated both in the case of $T_{DPDu}$ and $T_{SDDu}$ mainly in their preliminary versions.

\begin{table}
 \centering
  \caption{Percentages of the various cases of sunspot group data in tilt angle data bases $T_{MW}$ (1961-67) and $T_{SDDu}$ (1996-2010).}
  \begin{tabular}{lll}
  \hline
   Data base     &  $T_{MW}\&T_{MWDPD}$   &       $T_{SDD}$ \\

 \hline
 Number of cases & 506 & 108523 \\
  \hline
 Bipolar sunspot group & 55.93\% & 74.26\% \\
 Unipolar spots & 26.09\% & 25.74\% \\
 Only one portion measured & 13.83\% & \\
 Mixed-up nearby groups & 4.15\% &   \\
\hline
\end{tabular}
\end{table}

We have examined how the tilt angles of bipolar and unipolar (physically unipolar or incorrectly divided) groups depend on the separation of leading and following portions ($S_{MW}$). Figures 1-2 show the distribution of tilt angles $T_{MW}$ versus $S_{MW}$ in the case of bipolar and unipolar sunspot groups of Table 1, respectively.

\begin{figure}
\centering
\includegraphics[width=7cm]{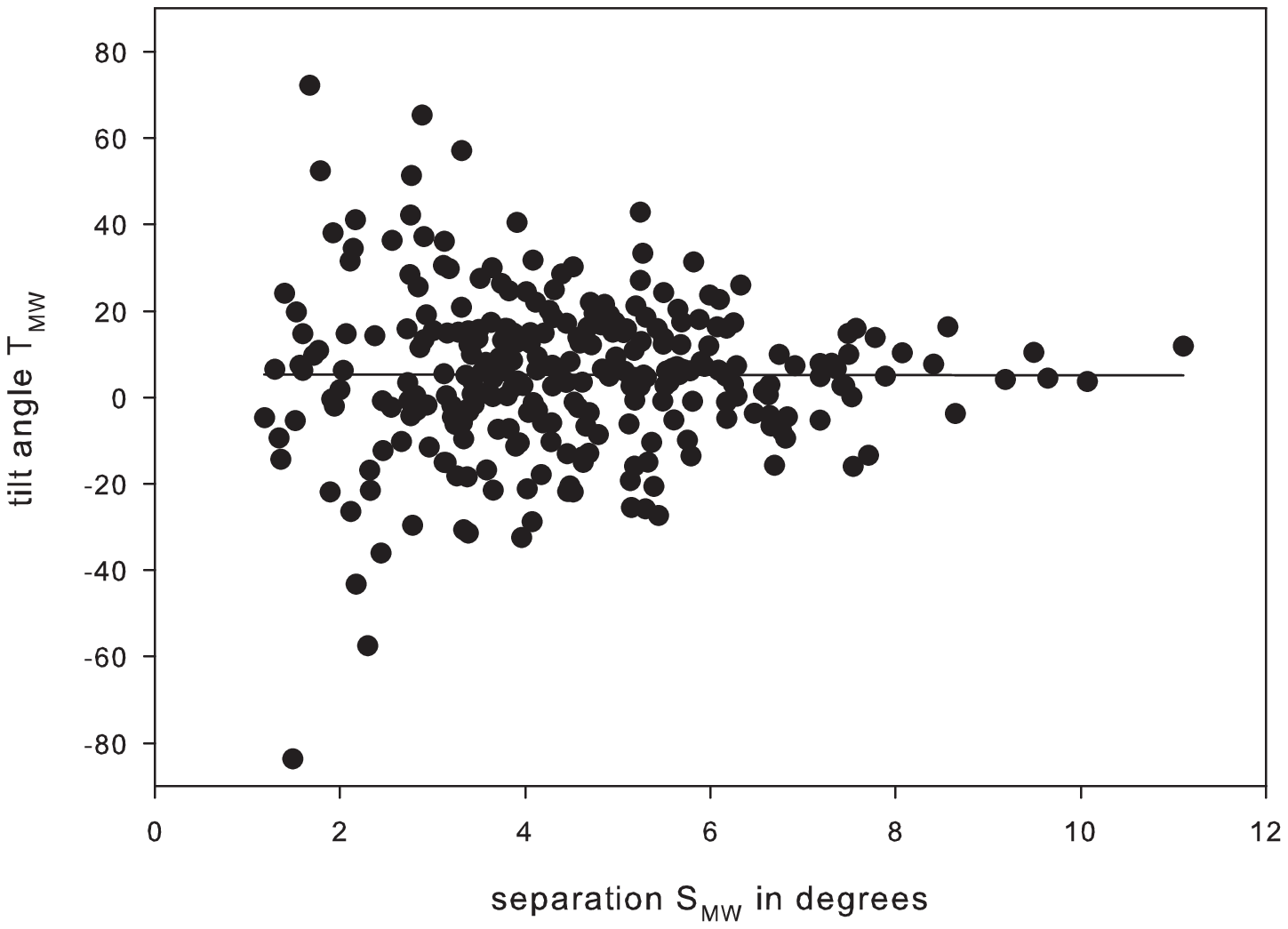}
 \caption{Tilt angle $T_{MW}$ versus separation $S_{MW}$ in the case of bipolar groups. The equation of the regression line is \newline $T_{MW}= 5.43(\pm2.95)-0.31(\pm0.62)S_{MW}$.}
 \label{fig1}
\end{figure}

\begin{figure}
\centering
\includegraphics[width=7cm]{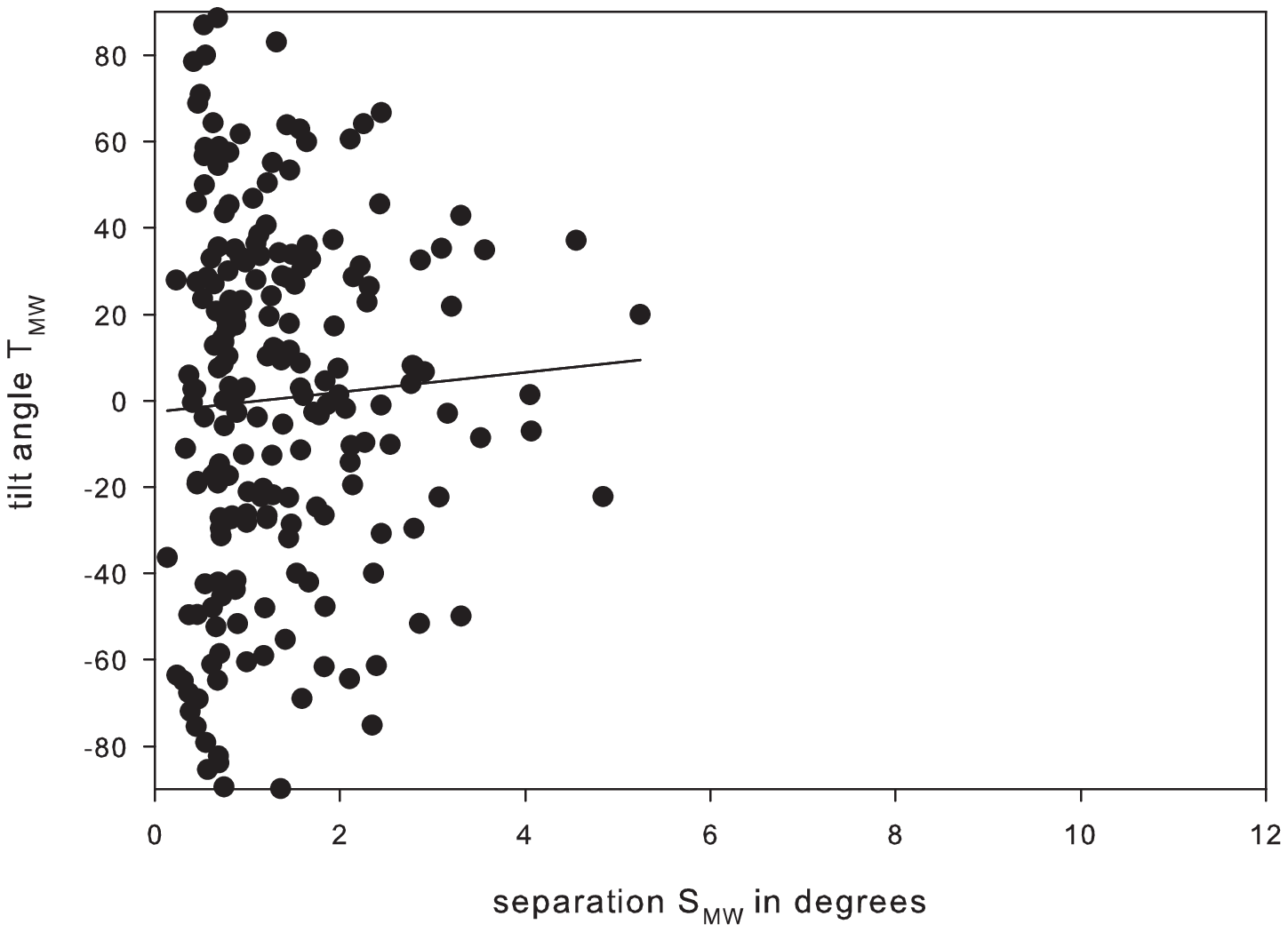}
 \caption{Tilt angle $T_{MW}$ versus separation $S_{MW}$ in the case of unipolar groups. The equation of the regression line is \newline $T_{MW}= -2.53(\pm5.09)+2.29(\pm3.12)S_{MW}$. }
 \label{fig2}
\end{figure}

It can be seen in these figures that the tilt angles of bipolar groups show a smaller dependence on $S_{MW}$ than those of the unipolar groups. The mean tilt angle of the unipolar groups is much smaller (0.54$\pm$2.86) than that of the bipolar groups (5.29$\pm$1.07). This means that the mean tilt angle of the whole data set depends on the ratio of bipolar and unipolar sunspot groups involved in the measurements.

\section{Filtering out divided groups}

About 1/3 of the effect of unipolar groups comes from the divided groups of $MW/KK$, and the decrease of their contribution can improve the statistical results. Thus, we looked for a suitable filter that can be used without a detailed investigation of sunspot groups.  We tested how the divided groups in the $MW$ (and $KK$) data set can be identified if we try to match them to the group of closest position in the $GPR$ (until 1976) or $DPD$ (after 1974) catalogue. In a simple case, $GPR/DPD$ contains only one group at about that heliographic position where $MW$ (or $KK$) contains two groups (one for the leading portion and one for the following portion). In this case, both groups of $MW$ ($KK$) can be only matched to the same group of $GPR/DPD$. About 10\% of the $MW/KK$ cases was matched twice or more times sharing the same pair in $GPR/DPD$. Sorting out these double or multiple pairs increases the mean tilt angle as it can be expected from the study of the pilot data base. For example, the mean of $T_{MW}$ in the interval 1917-1976 is $4.16\pm0.19$ without using this filter and it is $4.69\pm0.20$ with this filter. In 1974-1985, these values are $4.56\pm0.46$ and $5.00\pm0.47$ respectively. The mean $T_{DPDu}$ in this interval is $5.12\pm0.46$; and the mean $T_{MW}$ is closer to that of $T_{DPDu}$ if we use this filter. However, the tests show that the method is not perfect because there are cases when the divided groups are not identified. For example, this method does not work well if one of the portions of the divided group is omitted from the $MW/KK$ data base e.g. because one of the portions is at $|LCM|>60^{\circ}$ while another portion is at $|LCM|<60^{\circ}$. The time difference between the data bases and the nearby groups involved only in $GPR/DPD$ also decreases the efficiency of the matching process. This filter can be useful if it is important to involve all the available data into a study. If there is a need to improve the sample in the range of larger separation, the correct tilt angle can be also calculated by matching the leading and following portions to each other. 

\section{Filtering out unipolar groups}

If the unambiguity of the data has a higher importance than the size of the sample, a more robust filter can be usually more useful. Such a filter can be derived from Figures 1-2 by comparing the separation of bipolar and unipolar groups. It can be seen that the ratio of bipolar groups increases with increasing separation. The range of small separation (about $<2^{\circ}$) is dominated by unipolar groups while bipolar groups have usually a larger  ($>3^{\circ}$) separation. This provides a tool to decrease the effect of unipolar groups and indefinite cases. 

We investigated which criterion for separation is suitable to filter out unipolar groups. The statistical study of the pilot data set shows that the suitable threshold can be about $2.5-3^{\circ}$. Below $2.5^{\circ}$, the unipolar groups clearly dominate in the sample. The ratio of bipolar and unipolar groups is $\sim$ 1:1 in the interval $2-2.5^{\circ}$ and it is $\sim$3:1 in the interval $2.5-3^{\circ}$. If there is a need to keep the largest number of bipolar groups in the studied sample, the criterion of $S>2.5^{\circ}$ can be used. In this case, 84\% of the excluded groups is unipolar while 90\% of unipolar groups and 12\% of bipolar groups are excluded. If there is a need to exclude more unipolar groups, the criterion of $S>3^{\circ}$ can be used. In this case, 77\% of the excluded groups is unipolar while 94\% of unipolar groups and 19\% of bipolar groups are excluded. 

We have similar results if we use the larger sample of $T_{SDDu}$ (108523 cases  within the range of $|LCM|< 60^{\circ}$): $\sim$ 90\% of unipolar groups and 20\% of bipolar groups have smaller separation than $3^{\circ}$. The type of bipolar groups can be various in this range (e.g. small group with a few small spots at its maximal development; small developing group (often with small spots in a ring like structure); large old spot with a few small spots of opposite polarities somewhere around it; small complex group with a few umbrae of opposite polarities within a common penumbra). These types of groups cause that the scatter of the tilt data of bipolar groups with small separation is large. The criterion of $S>3^{\circ}$ filters out not only the majority of unipolar groups but this inhomogeneous subset of bipolar groups too.

\section{Differences between methods}

\begin{figure*}
\centering
\includegraphics[width=14cm]{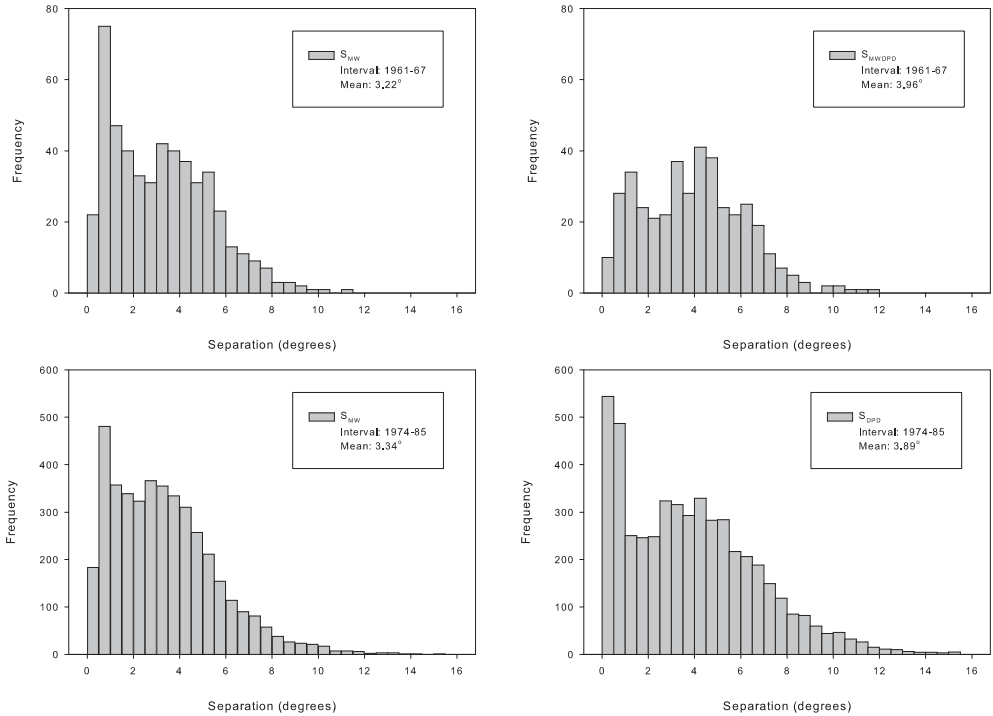}
 \caption{Frequency of the separation of leading and following portions of sunspot groups in various tilt angle data sets. panels: Left separation of sunspot groups based on the $MW$ data set ($S_{MW}$) in two time intervals. Right panels: separation of sunspot groups based on the $DPDu$ data set in the same two intervals ($S_{MWDPD}$ in 1961-67 and $S_{DPDu}$ in 1974-85).}
\end{figure*}

This section discusses those methodological differences that can be responsible for the differences and hidden sources of errors in tilt angle data. In Figure 3, we compared the histograms of the separations in two subsets of $MW$ and $DPD$ to reveal the differences between them.  In the upper panels, the histograms refer to the same sunspot groups measured with two different methods in 1961-67. The lower panels show the frequency of the tilt angles in the data set $T_{MW}$ the years 1974-85 and that of $T_{DPDu}$ on the same days within $|LCM|< 60^{\circ}$ but without matching them to each other. 
The histograms show that there are quite large differences between the frequencies of the separation in the data sets. The reason of these differences is the difference between the measuring methods.

At $MW$, a pore was measured in a similar way as an umbra. In this way, the calculation of tilt angles is based on both pores and umbrae. However, the $DPD$ (the official continuation of $GPR$) has to handle the pores and umbrae as $GPR$ did.  In the $GPR$, in the case of the pores or spots of area below 5 msh (millionths of the solar hemisphere) usually the whole spot area was measured and in the column of umbral area was zero.
The method of $DPD$ is developed to result in similar umbra and whole spot area to that of $GPR$. Each solar feature is classified as a "spot" which is darker than the photosphere and its whole spot area is measured. The umbra area is only measured in those cases when the software identifies one or more darker parts in the spot. Since the tilt angle $T_{DPDu}$ (and $T_{MWDPD}$) is based on the data in the column of umbral area, the pores are not included into the umbral tilt angle at all.  

The different umbra definition or separation/union of nearby umbrae in $MW$ and $DPD$ may also cause deviations. For example, there are groups with $T_{MW}$ but without $T_{DPDu}$ (and vice versa). 

The above-mentioned things may cause differences in the number of tilt angles and in the separations, which can be efficiently decreased by filtering out the groups of small separation as it can be seen in Figure 3. 

We have found that in the half of the cases when there is $T_{MW}$ but no $T_{DPDu}$, the tilt angle $T_{DPDws}$ can be in a good agreement with $T_{MW}$. 
Thus, it is useful to investigate the $T_{DPDws}$ tilt angles too.
The upper panel of Figure 4 shows the histogram of the separation derived from the $T_{DPDws}$ data. It can be seen that the relative ratio of the groups in the range of small separation $S_{DPDu}<2$ to the other groups is smaller here than it is in the lower right panel of Figure 3. This means that the majority of the tilt angles $T_{DPDu}$ with $S_{DPDu}<2$ comes from the groups which consist of one single spot containing more than one umbra. These cases can be also filtered out by using a threshold for $|T_{DPDu}-T_{DPDws}|$ as it can be seen in the lower panel of Figure 4. This filter allows selecting the most unambiguous cases when the tilt angles derived with two different methods are in a good agreement with each other.

\begin{figure}
\centering
\includegraphics[width=7cm]{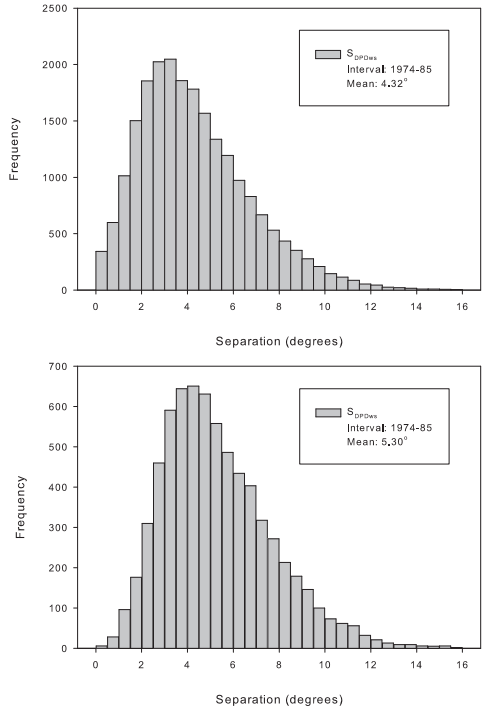}
 \caption{Frequency of the separation of leading and following portions of sunspot groups based on the $DPDws$ tilt angle data. Upper panel: $S_{DPDws}$ in the same time interval as in the lower panel of Figure 3. Lower panel: the same as in the upper panel in the case of $|T_{DPDu}-T_{DPDws}|<10^{\circ}$.}
\end{figure}

\begin{figure}
\centering
\includegraphics[width=7cm]{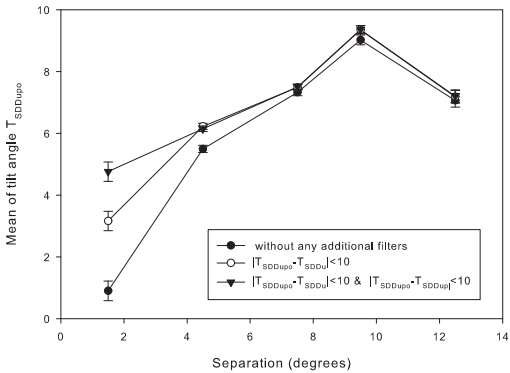}
 \caption{Mean of $T_{SDDupo}$ versus separation of sunspot groups $S_{SDDupo}$ calculated for separation bins of $3^{\circ}$ by using various selection criteria.}
\end{figure}

In Figure 5, it can be seen why the differences in the separation of involved groups are important from the point of view of mean tilt angles. The effect of the increase of mean separation on the mean tilt angle and the effect of various filters can be estimated based on tilt angle $T_{SDDupo}$ where the  information on umbral polarity is taken into account in the calculation.  It can be seen that the mean tilt angle is increasing with increasing separation that is in agreement with the result of \citet{H93}. This means that the mean tilt angle depends on the frequencies of the sunspot groups with various separations. It can be also seen that the mean tilt angle increases when we increase the number of more unambiguous cases by using some filters for the difference between the tilt angles calculated with different methods. The figure shows that these filters also affect the mean tilt angle of the selected sample in a different extent.  

\section{Tilt angles}

By using the possible filters above, we can investigate whether there are any significant systematic deviations between the tilt angles of $MW$ and $DPD$ data bases. To determine the cross-calibration factors, we follow the method of OLS-mean \citep{BF92} in the same way as it was described in the Section "Method of analysis" of \citet{BKC13} and similar to the method of \citet{Bal09}. Ordinary least-squares (OLS) linear regression model without an intercept term (i.e. the slope of a linear regression forced to pass through the origin) is applied to the data of two different observatories in two iterations. The slope derived in the first iteration is taken to be the initial estimate for a second analysis, and the standard error of the estimate ($\sigma_{fit}$) is determined. In the second analysis, the outliers are excluded when only points within $3*\sigma_{fit}$ from the first fit are taken. The OLS-mean method treats the variables symmetrically. Thus, we apply an inverse linear regression with the same iterations, switching the dependent variable with the independent variable. Finally, the arithmetic mean of the two OLS slopes is calculated as an estimate of the slope of the regression line. 

For the bipolar groups matched in 1961-67, we used the selection criteria of both $S_{MW}$ and $S_{MWDPD}$ larger than 3 degrees. In these cases, there is a large probability that the same parts of the group were measured in a similar way. The result of the comparison shows that there is no systematic deviation between the two data bases because the correction factor practically does not differ from one. 
 
\begin{equation}
   T_{MWDPD}=1.02(\pm 0.02) *T_{MW}.
\end{equation}

For the years 1977-85, the correction factor is also a negligible value determined when we use the same selection criteria and unambiguous pairs of groups of $DPD$ and $MW$.
\begin{equation}
   T_{DPDu}=1.01(\pm 0.01) *T_{MW}.
\end{equation}

Equations (3) and (4) show that the difference between the measuring methods does not result in a systematic difference between the tilt angles in the unambiguous cases.

\begin{table}
 \centering
  \caption{Statistics of the differences between sunspot group data in tilt angle data bases $MWDPD$ and $MW$. The studied sample consists of 206 cases when both $T_{MWDPD}$ and $T_{MW}$ are determined and both $S_{MW}$ and $S_{DPD}$ are larger than $3^{\circ}$.}
  \begin{tabular}{lll}
  \hline
   $|T_{MWDPD} - T_{MW}|$     &  Number of cases   &       Percentage \\
  \hline
 $<5^{\circ}$ & 164 & 79\% \\
 $<10^{\circ}$  & 194 & 94\% \\
 $<15^{\circ}$ & 201 & 97\%\\
\hline
\end{tabular}
\end{table}

The random deviations are not too large either. Table 2 shows the statistics of the differences between $T_{MWDPD}$ and $T_{MW}$. (Without the criteria of $S_{DPDu}>3$ and $S_{MW}>3$, the percentages are  6-9\% smaller.) Our test shows that the cause of the large deviation is usually the difference between the longitudes of the area-weighted centroids in $MW$ and $DPD$. In these cases, the difference between $T_{MWDPD}$ and $T_{MW}$ mainly depends on the area data of the umbrae which determine the position of the longitude separating the leading and following portions. A small difference in the separating longitude may change the assignment of some umbrae in the middle of the group to the leading or to the following portion. This shows that some differences between the area measurements can cause deviations of tilt angles.

\section{Area of sunspot groups}

Because of the important role of area data, it is useful to compare the pairs of area values belonging to the same groups in the various data bases. By using the same OLS-mean method described in the previous section, we compute the correction factor between the area $A_{MW}$ published in the MW tilt angle data base  and corrected umbral area $A_{MWDPD}$ derived by the software of $DPD$ from the scans of MW Directs. The time interval 1961--67 yielded a sample of 537 pairs of area data. We also compared these data with the corrected umbral area in $GPR$ ($A_{GPRu}$). 
The equations of the linear regression are:
\begin{equation}
   A_{GPRu}=1.84(\pm 0.06) *A_{MW}.
\end{equation}

\begin{equation}
   A_{MWDPD}=1.98(\pm 0.05) *A_{MW} 
\end{equation}

\begin{equation}
 A_{GPRu}=0.95(\pm 0.02) *A_{MWDPD}.
\end{equation}

Equation (7) shows that there is no substantial systematic difference between the $GPR$ and $MWDPD$ area data. This is in good agreement with the results of \citep{Bea01, BKC13} claiming that the correction factor between the $GPR$ and $DPD$ area data is practically equal to 1. However, equations (5) and (6) show that there is a quite large correction factor between the $A_{MWDPD}$ and $A_{MW}$ data although both of them were measured in the same observation. This supports the arguments of \citet{F14} who concluded that the large deviation of $A_{MW}$ data from the $GPR$ area data comes from the measuring method of the $MW$ data. 

To investigate the relationship on a larger sample, we compute the correction factor between the area data for the overlapping intervals of $DPD$ and $MW$. The number of the data pairs $A_{MW}$ and $A_{DPDu}$ is 4644 in the interval 1974-85. The result is only slightly different from equation (6):

\begin{equation}
   A_{DPDu}=1.91(\pm 0.03) *A_{MW}.
\end{equation}
 The reason of the small difference can be caused by the difference in the size of samples as well as a small time dependent variation of data which can be expected in the case of any data bases  \citep{BKC13}  The comparison can be made for the overlapping intervals of $DPD$ and $KK$ data too. The number of the data pairs $A_{KK}$ and $A_{DPDu}$ is 4995 in the interval 1974-87. The result is:
\begin{equation}
   A_{DPDu}=1.71(\pm 0.02) *A_{KK} 
\end{equation}

These correction factors have to be used when area data of both $DPD$ and $MW/KK$ are used in the same study. 

\section{Joy's law}

We examine how the selection criteria described above affect the slope of Joy's law. Figures 6-7 show the effect of filters $S_{MW}>3$ and $S_{KK}>3$ on the Joy's law based on the $MW$ and $KK$ data, respectively. It can be seen that the mean tilt angles calculated in bins of $5^{\circ}$ increase when the filter is applied and the increase is larger at larger latitudes. The slopes of the regression lines fitted in the latitudinal range of $0-30^{\circ}$ in the various cases are listed in Table 3, which shows that the slopes determined by using the group with $S>3$ are in a better agreement with each other than the original ones. These values have also a smaller deviation from the slope calculated from the data series of the Pulkovo data base (CSA) data ($0.38 \pm 0.03$) which is also based on white-light observations \citep{I12}.  In Table 3 we indicated the slopes of regression lines that are forced through the origin so that the comparison could be clearer. However, it is an open question whether regression lines have to be forced through the origin. Some authors include constant in equations (e.g. \citealt{LU12, MN13}) while some other authors determine the Joy's law with no intercept (e.g. \citealt{I12, Dasi10}). 

\begin{figure}
\centering
\includegraphics[width=7cm]{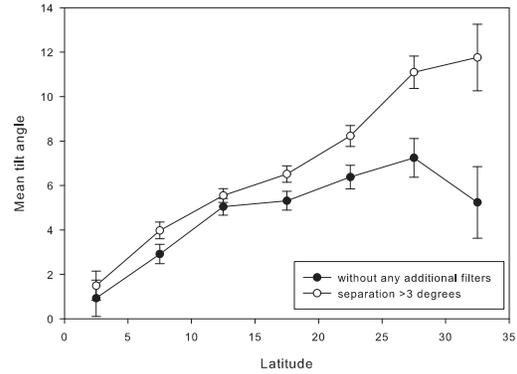}
 \caption{Joy's law derived from $T_{MW}$ with and without using the selection criterion $S_{MW}>3^{\circ}$. Error bars represent $\pm$ one standard error of the mean in all the figures. The latitude is $|B|$.}
 \label{fig6}
\end{figure}

\begin{figure}
\centering
\includegraphics[width=7cm]{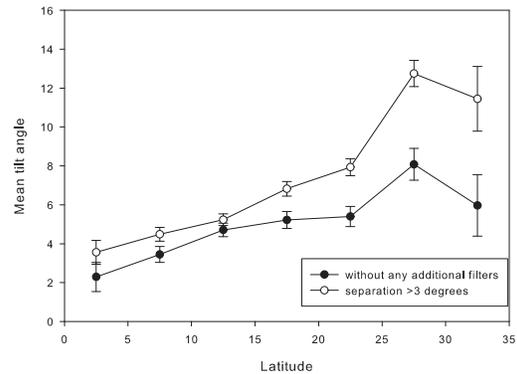}
 \caption{Joy's law derived from $T_{KK}$ with and without using the selection criterion $S_{KK}>3^{\circ}$. }
 \label{fig7}
\end{figure}

\begin{table}
\centering
 \caption{Slope of the Joy's law calculated for the whole interval of the various tilt angle data bases in the latitude range of $\pm 30^{\circ}$. (The regression line is forced through the origin.)}
 \begin{tabular}{lll}
  \hline
  Data base & Slope      &      Slope if $S>3$ \\
  \hline
 $T_{MW}$ &  0.292($\pm 0.019$)  &  0.396($\pm 0.016$)  \\
 $T_{KK}$  &  0.294($\pm 0.027$)  &  0.423($\pm 0.037$)  \\
 $T_{DPDu}$ &   0.327($\pm 0.017$) &  0.422($\pm 0.016$) \\ 
 $T_{SDDu}$ &   0.373($\pm 0.042$) &  0.423($\pm 0.025$) \\ 
\hline
\end{tabular}
\end{table}

We have investigated whether the $DPD$ clarifies the question of intercept of the regression line after selecting out the most ambiguous cases with two filters of $S_{DPDu}>3^{\circ}$ and $|T_{DPDu}-T_{DPDws}|<10$ applied for each solar cycle. In Figure 8, we can see that in Cycles 22-23 the constant of the regression line is close to zero ($0.38|B|+0.09$ and $0.43|B|-0.05$, respectively) but it is somewhat larger in Cycle 21 ($0.29|B|+2.29$). Except for this and a few other small differences, the curves are quite similar to each other. 

The most intriguing feature of these diagrams is that the rate of tilt angle increase toward the poles is not constant. There is a well recognizable plateau at about the latitude of $15^{\circ}$. The close spatial connection with the mean latitude of the sunspot groups ($15.8^{\circ}$, Std. Dev. $7.4^{\circ}$) may imply that the plateau has a physical connection with the main part of the active region belt. We have checked whether the comparison of the hemispheres reveals further details. It can be seen in Figure 9 that the plateau is expressed in each cycle on the northern hemisphere and it is recognizable in the southern hemisphere though less unambiguously. Despite this difference, the slope of the regression line fitted in the latitudinal range of $0-30^{\circ}$ is about the same in each cycle and in both hemispheres except for the southern hemisphere in Cycle 21 as it can be seen in Table 4.

We have examined whether the plateau arises due to the fact that the $T_{DPDu}$ is calculated with no polarity information. This can be checked by using $T_{SDDupo}$ data. We also exploited that it has a large enough sample to decrease the latitudinal bins from $5^{\circ}$ to $3^{\circ}$
to get a more detailed curve. We plotted the median values so that it could be excluded that it is only an effect on means. It can be seen in Figure 10 that the diagram of $T_{SDDupo}$ contains a well recognizable plateau in the southern hemisphere and somewhat less expressed one in the northern hemisphere. This implies that the lack of the polarity information is not the reason of non-linearity. 

Figure 5 may imply that the groups with separation larger than about $10^{\circ}$ have a special effect on the mean tilt angle thus the large complex groups or long trains of inseparable nearby groups may be also mentioned as possible reasons of the plateau-like deviation. 
To reveal any effects of them on the Joy's law, we plotted the medians in three separation ranges in Figure 11. This figure shows that the curve of groups with separation of $S_{SDDupo}>10^{\circ}$ explicitly deviates from the two other ones but the plateau is not their result because it can be see well in the ranges of smaller separation. 
The time series of this latter subset of data was divided into several subintervals of the solar cycle to investigate the temporal variation of the Joy's law. Figure 12 shows that the plateau is recognizable in each subinterval although the median values have quite large differences because of the decreased sample size. It seems that the plateau is narrower in the years around the solar maximum than in the years of lower activity. 

The results show that the plateau may be detected in several suitably chosen subsets of data.
To provide much clearer evidence for it and to reveal the causes of the found deviations, further detailed studies will be needed. However, Figures 8-12 are suitable to demonstrate the advantages of the new tilt angle data sets. The large sample of tilt angles and various types of additional information allow studying some characteristics of tilt angles in details e. g. the fine structure of the Joy's law.

\begin{figure}
\centering
\includegraphics[width=7cm]{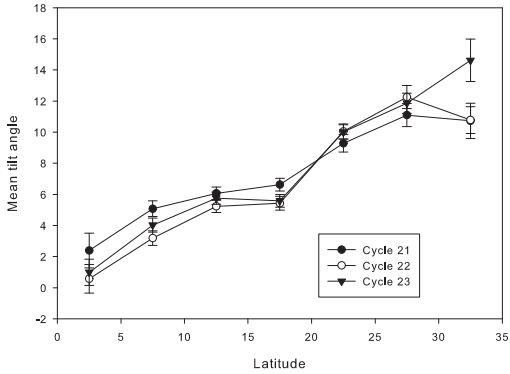}
 \caption{Joy's law derived from $T_{DPDu}$ by using the selection criteria $S_{DPDu}>3^{\circ}$ and $|T_{DPDu}-T_{DPDws}|<10^{\circ}$. }
 \label{fig8}
\end{figure}

\begin{figure}
\centering
\includegraphics[width=7cm]{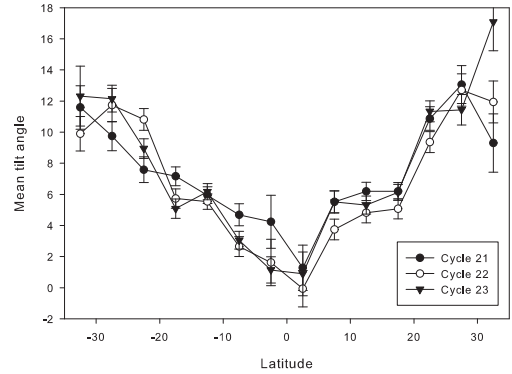}
 \caption{The same as Figure 7 but separated for the northern and southern hemispheres by keeping the sign of latitude.}
 \label{fig9}
\end{figure}

\begin{figure}
\centering
\includegraphics[width=7cm]{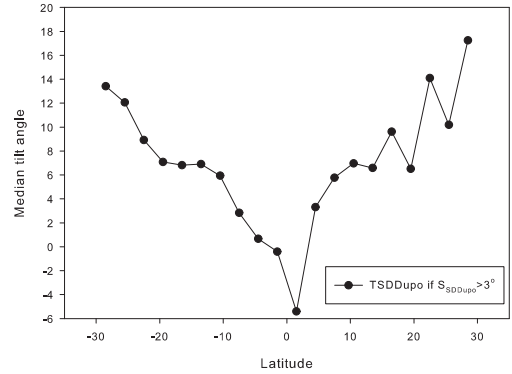}
 \caption{Joy's law derived from $T_{SDDupo}$ in latitudinal bins of $3^{\circ}$.}
 \label{fig10}
\end{figure}

\begin{table}
\centering
 \caption{Equation of Joy's law calculated for $T_{DPDu}$ in three solar cycles shown in Figure 9.}
 \begin{tabular}{cll}
  \hline
  C & Northern hemisphere &    Southern hemisphere \\
 \hline
 21 & $0.43(\pm.07)B+0.76(\pm1.13)$ & $-0.21(\pm.07)B+3.35(\pm0.37)$  \\
 22 & $0.46(\pm.06)B-1.00(\pm1.07)$  & $-0.43(\pm.05)B-0.10(\pm0.93)$  \\
 23 & $0.41(\pm.07)B+0.68(\pm1.27)$  & $-0.41(\pm.06)B-0.05(\pm1.03)$  \\
\hline
\end{tabular}
\end{table}

\begin{figure}
\centering
\includegraphics[width=7cm]{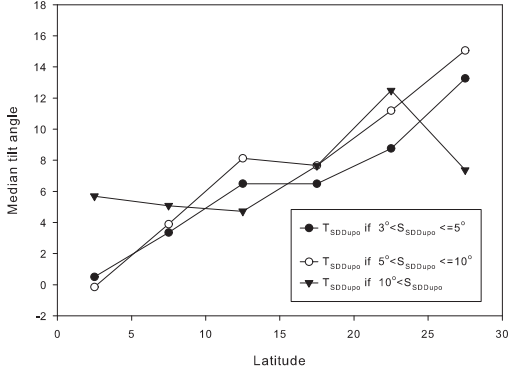}
 \caption{Joy's law derived from $T_{SDDupo}$ in three separation ranges  in latitudinal bins of $5^{\circ}$ of $|B|$.} 
 \label{fig11}
\end{figure}

\begin{figure}
\centering
\includegraphics[width=7cm]{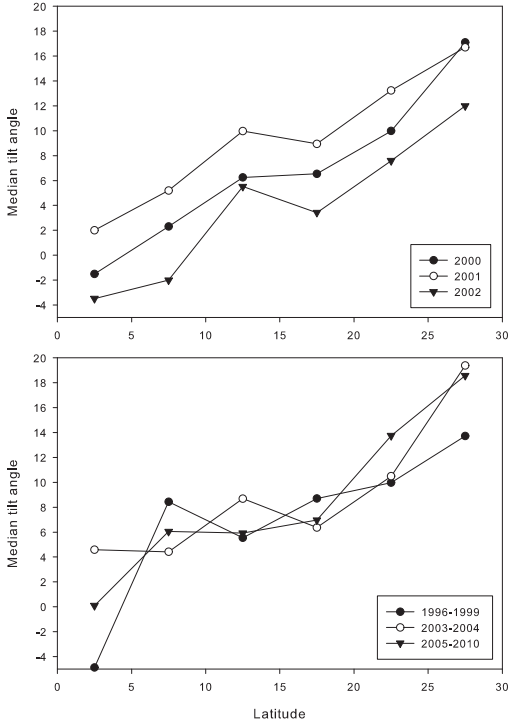}
 \caption{ Joy's law derived from $T_{SDDupo}$ for the groups of $S_{SDDupo}<10^{\circ}$. The time interval is divided into six parts to see the temporal variation. Upper panel: years around solar maximum. Lower panel: years before and after maximum years.}
 \label{fig12}
\end{figure}

\section{Summary}

We have compared the traditional tilt angle data bases of Mount Wilson ($MW$), Kodaikanal ($KK$) observatories with the recently published data bases ($DPD$, $SDD$) of Debrecen Observatory in different time intervals by statistical methods and case studies. New tilt angles have been derived from $MW$ white-light observations for 1961--67 by using the software of $DPD$. We have compared the positions of the leading and following portions of the $MW$ tilt angle data base with the positions in the photographic images and with the polarity information on that part of the group in the $MW$ polarity drawings. Based on these data, we have revealed several differences between the data bases coming from the different measuring methods. We have tested a few new methods to filter out the majority of incorrect tilt angles and to decrease the deviations because of random errors. We have calculated the calibration factors between umbral area of $MW$, $KK$, $GPR$ and $DPD$. Our results show that there is no systematic deviation between the traditional and new tilt angle data sets. The filters and calibrations enhance the reliability and homogeneity of the tilt angle data bases.

Because of these revisions, improved curves of the Joy's law have been obtained by applying various filters on the tilt angle data. Our results show that the applied filters increase the slope of the Joy's law; in addition, the slopes derived from different data sets get closer to each other.
The decrease of the unambiguous tilt data may allow recognizing new types of characteristics of tilt angles. 
The most unexpected result based on the new Debrecen data is that the curves representing the Joy's law exhibit an indication for a conspicuous feature: a plateau in the domain around the mean latitude of the active region belt at $\sim16^{\circ}$. This may imply a still unknown contributor to the fine structure of the Joy's law based on sunspot group tilt angles. Further research is required to provide much clearer evidence or a refutation of the possible significance of this feature but the advantages of the new data may be seen well.

\section*{Acknowledgments}

I thank N. Krivova, S. Solanki, and M. Dasi-Espuig for initiating the study of DPD tilt angles and for helpful discussions on the data. I thank A. Ludm\'any and Y.-M. Wang for improving the manuscript.  I am grateful to the referee for suggesting some extension of the study.
This study includes data from the synoptic program at the 150-Foot Solar Tower of the Mt. Wilson Observatory. The Mt. Wilson 150-Foot Solar Tower is operated by UCLA, with funding from NASA, ONR and NSF, under agreement with the Mt. Wilson Institute. Mt. Wilson Solar Photographic Archive Digitization Project is based upon work supported by the National Science Foundation under Grant No.0236682. SOHO is a project of international cooperation between ESA and NASA. We wish to acknowledge the several generations of observers at the various observatories and institutes who took the ground-based observations covering many decades and made them public. The research leading to these results has received funding from the European Community's Seventh Framework Programme (FP7/2007-2013) under grant agreement eHEROES (project No. 284461).

\label{lastpage}

\end{document}